\documentclass[a4paper, oneside, twocolumn, notitlepage, 10pt]{extarticle_ecoc}
\usepackage[nolist]{acronym}
\usepackage{tikz}
\usepackage{pgfplots}
\usetikzlibrary{shapes,arrows,positioning,calc, matrix,spy,patterns}
\usetikzlibrary{decorations.pathreplacing,calligraphy}
\usetikzlibrary{colorbrewer}
\usetikzlibrary{fit}
\usepgfplotslibrary{colorbrewer}
\usepackage{subcaption}
\usepackage{booktabs}
\usepackage{ecoc}
\usetikzlibrary{external}
\definecolor{kit-green100}{rgb}{0,.59,.51}
\definecolor{kit-green70}{rgb}{.3,.71,.65}
\definecolor{kit-green50}{rgb}{.50,.79,.75}
\definecolor{kit-green30}{rgb}{.69,.87,.85}
\definecolor{kit-green15}{rgb}{.85,.93,.93}
\definecolor{KITgreen}{rgb}{0,.59,.51}

\definecolor{KITpalegreen}{RGB}{130,190,60}
\colorlet{kit-maigreen100}{KITpalegreen}
\colorlet{kit-maigreen70}{KITpalegreen!70}
\colorlet{kit-maigreen50}{KITpalegreen!50}
\colorlet{kit-maigreen30}{KITpalegreen!30}
\colorlet{kit-maigreen15}{KITpalegreen!15}

\definecolor{KITblue}{rgb}{.27,.39,.66}
\definecolor{kit-blue100}{rgb}{.27,.39,.67}
\definecolor{kit-blue70}{rgb}{.49,.57,.76}
\definecolor{kit-blue50}{rgb}{.64,.69,.83}
\definecolor{kit-blue30}{rgb}{.78,.82,.9}
\definecolor{kit-blue15}{rgb}{.89,.91,.95}

\definecolor{KITyellow}{rgb}{.98,.89,0}
\definecolor{kit-yellow100}{cmyk}{0,.05,1,0}
\definecolor{kit-yellow70}{cmyk}{0,.035,.7,0}
\definecolor{kit-yellow50}{cmyk}{0,.025,.5,0}
\definecolor{kit-yellow30}{cmyk}{0,.015,.3,0}
\definecolor{kit-yellow15}{cmyk}{0,.0075,.15,0}

\definecolor{KITorange}{rgb}{.87,.60,.10}
\definecolor{kit-orange100}{cmyk}{0,.45,1,0}
\definecolor{kit-orange70}{cmyk}{0,.315,.7,0}
\definecolor{kit-orange50}{cmyk}{0,.225,.5,0}
\definecolor{kit-orange30}{cmyk}{0,.135,.3,0}
\definecolor{kit-orange15}{cmyk}{0,.0675,.15,0}

\definecolor{KITred}{rgb}{.63,.13,.13}
\definecolor{kit-red100}{cmyk}{.25,1,1,0}
\definecolor{kit-red70}{cmyk}{.175,.7,.7,0}
\definecolor{kit-red50}{cmyk}{.125,.5,.5,0}
\definecolor{kit-red30}{cmyk}{.075,.3,.3,0}
\definecolor{kit-red15}{cmyk}{.0375,.15,.15,0}

\definecolor{KITpurple}{RGB}{160,0,120}
\colorlet{kit-purple100}{KITpurple}
\colorlet{kit-purple70}{KITpurple!70}
\colorlet{kit-purple50}{KITpurple!50}
\colorlet{kit-purple30}{KITpurple!30}
\colorlet{kit-purple15}{KITpurple!15}

\definecolor{KITcyanblue}{RGB}{80,170,230}
\colorlet{kit-cyanblue100}{KITcyanblue}
\colorlet{kit-cyanblue70}{KITcyanblue!70}
\colorlet{kit-cyanblue50}{KITcyanblue!50}
\colorlet{kit-cyanblue30}{KITcyanblue!30}
\colorlet{kit-cyanblue15}{KITcyanblue!15}

\definecolor{KITbraun}{RGB}{167,130,46}

\definecolor{cb-1}{HTML}{4477AA}
\definecolor{cb-2}{HTML}{EE6677}
\definecolor{cb-3}{HTML}{228833}
\definecolor{cb-4}{HTML}{CCBB44}
\definecolor{cb-5}{HTML}{66CCEE}
\definecolor{cb-6}{HTML}{AA3377}
\definecolor{cb-7}{HTML}{BBBBBB}

\pgfplotscreateplotcyclelist{cb list}{
	cb-1,cb-2,cb-3,cb-4,cb-5,cb-6,cb-7
}

\tikzstyle{surroundDerivation} = [rectangle, rounded corners, inner sep=0.05cm, dashed, very thick]

\usepackage{colortbl}
\usepackage{xcolor}
\usepackage{pgfmath} %

\begin{document}
\selectlanguage{english}    %

\begin{acronym}[TROLL]
    \acro{ASE}[ASE]{amplified spontaneous emission}
	\acro{AWGN}[AWGN]{additive white Gaussian noise}
    \acro{CCDF}[CCDF]{complementary cumulative distribution function}
    \acro{CD}[CD]{chromatic dispersion}
    \acro{CDC}[CDC]{chromatic dispersion compensation}
    \acro{DSP}[DSP]{digital signal processing}
	\acro{EEPN}[EEPN]{equalization-enhanced phase noise}
    \acro{FDPE}[FDPE]{frequency-dependent phase error}
    \acro{FDPN}[FDPN]{frequency-dependent phase noise}
    \acro{GN}[GN]{Gaussian noise}
    \acro{IDR}[IDR]{ideal data remodulation}
	\acro{LO}[LO]{local oscillator}
    \acro{NL}[NL]{nonlinearities}
    \acro{PC}{phase compensation}
    \acro{SNR}[SNR]{signal-to-noise ratio}
    \acro{TGN}[TGN]{temporal Gaussian noise}
    \acro{TRx}[TRx]{transceiver}
\end{acronym}

\title{Modeling and Mitigation of\\Equalization-Enhanced Phase Noise}%

\author{
	Benedikt Geiger\textsuperscript{(1)}, Fred Buchali\textsuperscript{(2)}, Vahid Aref\textsuperscript{(2)}, and Laurent Schmalen\textsuperscript{(1)}
}

\maketitle                  %

\begin{strip}
    \begin{author_descr}
    
       \textsuperscript{(1)} Communications Engineering Lab (CEL), Karlsruhe Institute of Technology (KIT), \\ \hspace*{2.25ex}Hertzstraße 16, 76187 Karlsruhe, Germany,
       \textcolor{blue}{\uline{benedikt.geiger@kit.edu}}
    
       \textsuperscript{(2)} Nokia, Magirusstr. 8, 70469 Stuttgart, Germany
    \end{author_descr}
\end{strip}

\def\e{\mathrm{e}}
\def\j{\mathrm{j}}

\renewcommand\footnotemark{}
\renewcommand\footnoterule{}

\begin{strip}
    \begin{ecoc_abstract}
\Ac{EEPN} emerges as a key performance limitation in high symbol-rate coherent transmission systems. In this paper, we highlight recent advances in modeling \ac{EEPN} and show that the temporal Gaussian noise model reproduces the characteristic burst-like SNR degradation, enabling efficient system simulation.
\textcopyright2026 The Author(s)
    \end{ecoc_abstract}
\end{strip}

\tikzset{lsblock/.style = {rectangle, thick, draw, minimum width=1.4cm, minimum height=0.8cm, rounded corners=1.6mm, font=\footnotesize},}
\newcommand{\mulrel}[2]{
	\node [draw, thick, circle, minimum size = 0.3cm, #2] (#1) {};
	\draw [] (#1.south east) -- (#1.north west);
	\draw [] (#1.south west) -- (#1.north east);	
}

\begin{figure*}[!b]
    \vspace{-0.4cm}
	\begin{center}
		\resizebox{0.9\textwidth}{!}{%
			\tikzsetnextfilename{block_diagram}
			\includegraphics{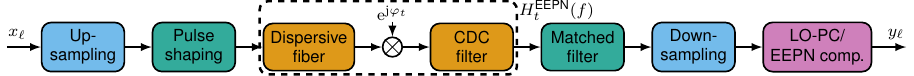}
		}
	\end{center}
    \vspace{-0.3cm}
	\caption{Block diagram of the equivalent baseband model used to investigate \acs{EEPN}.}
	\label{fig:system_model}
\end{figure*}

\section{Introduction}
The continuously growing demand for network capacity has driven symbol rates in coherent optical communication systems beyond \SI{100}{GBd}~\cite{Nokia_ICEX_800G_2025}. In particular, the rapid expansion of artificial intelligence applications is accelerating the need for cost- and power-efficient pluggable transceivers~\cite{AI_pluggables}. However, in these systems, \ac{EEPN} emerges as a key performance limitation at such high symbol rates~\cite{xu_study_2024,xu_system_2023}.

\ac{EEPN} arises from the interplay between the \ac{LO} phase noise and the digital \ac{CDC} filter~\cite{shieh_equalization-enhanced_2008}. As symbol rates and accumulated dispersion increase, this effect becomes more pronounced, and effectively tightens the linewidth requirements of the \ac{LO}. A particular challenge of \ac{EEPN} is its burst-like behavior, which leads not only to severe \ac{SNR} degradations, but also to potentially uncorrectable block errors in forward error correction~\cite{Geiger25ECOC, xu_study_2024}.

Therefore, \ac{EEPN} has recently received increasing attention and has been studied from several perspectives, including modeling~\cite{Geiger25OFC,Geiger26OECC,Geiger25ECOC,peng_dynamic_channel_charaterization}, mitigation~\cite{qiu_mitigation_2024, jung_mitigating_2024,abolfathimomtaz_equalization-enhanced_2024, abolfathimomtaz_receiver_2025, zhu_overcoming_2025}, statistical characterization~\cite{xu_study_2024,martins_frequency-band_2024}, \ac{DSP}-aware analysis~\cite{balducci_modeling_2025,jung_equalization-enhanced_2025,Buchali26OFC, arnould_equalization_2019}, and system-level investigations~\cite{ye_phenomenological_2022, lavery_state_2025, sun_800g_2020}.

In this paper, we highlight recent advances in modeling \ac{EEPN}. We revisit the fundamental distortion mechanism underlying \ac{EEPN}, show how it can be described as a \ac{FDPE}, and analyze how it manifests in detail. Furthermore, we study the theoretical \ac{EEPN} mitigation gain of different receiver structures (\ac{DSP} blocks), characterized by the order of the \ac{FDPE} they compensate. Finally, we show that the \ac{TGN} model captures the burst-like behavior of \ac{EEPN}, and can also predict performance after \ac{EEPN} mitigation.

\section{System Model} %
Since other distortions have a negligible interaction with \ac{EEPN}, we consider a minimal system in which \ac{EEPN} is the only distortion, as shown in Fig.~\ref{fig:system_model}. First, modulation symbols $x_\ell$ are generated at a symbol rate of \mbox{$R = \SI{130}{GBd}$}, and upsampled by a factor of \num{2}. Next, a root-raised cosine filter with a roll-off factor of $\num{0.01}$ is applied. We model a terrestrial, nonlinearity-free transmission over an \mbox{$L = \SI{2850}{km}$} long Corning TXF fiber, having a chromatic dispersion of \mbox{$D = 23\,\text{ps}\,\text{nm}^{-1}\,\text{km}^{-1}$} at a wavelength of \mbox{$\lambda = \SI{1550}{\nano \meter}$}. At the coherent receiver, the phase noise $\varphi_t$ is added due to the finite linewidth \mbox{$\Delta \nu \! = \! \SI{115}{kHz}$} of the \ac{LO}, which is modeled as a Wiener process. In the digital domain, a \ac{CDC} and matched filter are applied. Then, the signal is downsampled, and an \ac{LO}-\ac{PC} reverses the instantaneous \ac{LO} phase walk-off $\varphi_{t=\ell/R}$~\cite{arnould_equalization_2019}. Since we apply the \ac{CDC} filter at the receiver, the transmitter phase noise passes through the fiber channel and the \ac{CDC} filter and does not result in \ac{EEPN}~\cite{shieh_equalization-enhanced_2008}.

\begin{figure*} [!t]
    \input{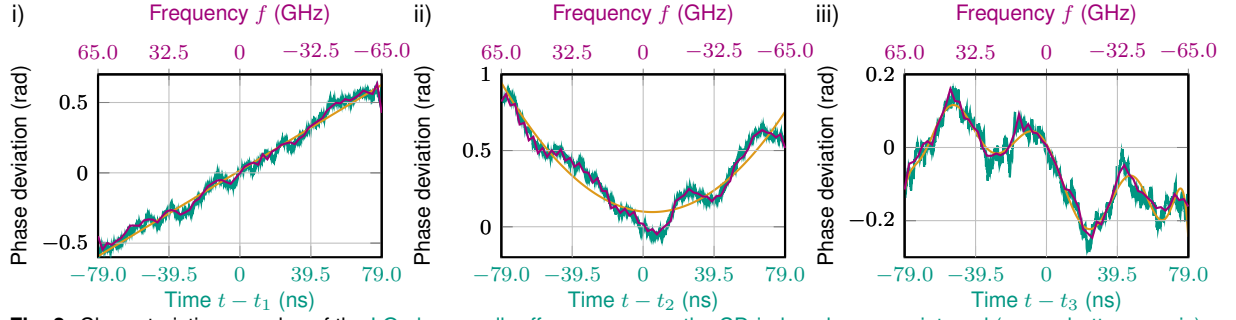}
    \vspace{-0.55cm}
    \caption{Characteristic examples of the \textcolor{KITgreen}{\ac{LO} phase walk-off $\varphi_t - \varphi_{t_i}$ over the \acs{CD}-induced memory interval (green, bottom x-axis)}, together with the \textcolor{KITpurple}{estimated \ac{FDPE} $\theta_{t_i}(f)$~\cite{Geiger25OFC} (purple, top x-axis)}, and the \textcolor{KITorange}{polynomial approximation (orange)}. The \ac{FDPE} manifests as timing offset in i), dispersive behavior in ii), and a superposition of higher-order distortions in iii).}
    \label{fig:phase_examples}
\end{figure*}

\section{EEPN as Frequency-Dependent Phase Noise}
To understand the fundamental nature of \ac{EEPN}, we revisit how the \ac{LO} phase noise interacts with the \ac{CDC}. The key observation is that the \ac{CDC} filter introduces a frequency-dependent group delay \mbox{$\tau_{\text{gr}}(f) = \frac{\lambda^2}{c}DLf$} not only to the signal of interest but also to the \ac{LO} phase, where $c$ is the speed of light~\cite{neves_enhanced_2021}.

Consequently, at a time instance $t_0$ and frequency $f$, the phase $\theta_{t_0}(f)$ after the \ac{CDC} filter is approximately given by~\cite{neves_enhanced_2021,Peng26OFC}
\begin{equation}
    \theta_{t_0}(f) = \varphi_{t_0 - \frac{\lambda^2}{c}DLf}.
    \label{eq:EEPN_phase}
\end{equation}
This means that each frequency component $f$ observes a delayed (or advanced) sample of the \ac{LO} phase noise $\varphi_t$ around $\varphi_{t_0}$. In particular, the phase samples observed across the signal bandwidth correspond to \ac{LO} phase samples within the \ac{CD}-induced memory $\tau_{\text{CD}} \approx \frac{\lambda^2}{c}DLR$. As a result, the phase noise becomes frequency-dependent after the \ac{CDC} filter, and can no longer be modeled as a common (frequency-independent) phase rotation. 

The \ac{FDPN} can be modeled by a time-varying all-pass filter with time-varying frequency response~\cite{Geiger25OFC,peng_dynamic_channel_charaterization,Peng26OFC}
\vspace{-0.15cm}
\begin{equation}
    H^{\text{EEPN}}_{t}(f) = \mathrm{exp} \left(\j \theta_{t}(f) \right),
    \label{eq:phase_error_translation}
    \vspace*{-0.15cm}
\end{equation}
where the time variation arises from the temporal evolution of the \ac{LO} phase walk-off.

Since the \ac{LO}-\ac{PC} removes only a frequency-independent phase term, the \ac{FDPN} results in a residual \ac{FDPE} after the \ac{LO}-\ac{PC}~\cite{Geiger25OFC,Peng26OFC}
\begin{equation}
    \theta^{\text{FDPE}}_{t}(f) = \theta_{t}(f) - \varphi_{t} \overset{(1)}{=} \varphi_{t - \frac{\lambda^2}{c}DLf} - \varphi_t.
\end{equation}
To analyze the distortions caused by the \ac{FDPE} in more detail, we approximate the \ac{FDPE}, or equivalently the phase error induced by the phase walk-off over time, by an $N$\textsuperscript{th}-order polynomial~\cite{Geiger25OFC}
\vspace{-0.22cm}
\begin{equation}
    \theta^{\text{FDPE}}_{t}(f) \approx \sum_{n=0}^N a^{(n)}_t \cdot f^n,
    \label{eq:EEPN:Höhere_Ordning_Phase}
    \vspace{-0.22cm}
\end{equation}
where the coefficients $a^{(n)}_t$ are chosen to minimize the mean squared error between the true \ac{FDPE} and its approximation.

Fig.~\ref{fig:phase_examples} shows the \ac{LO} phase walk-off over the \ac{CD}-induced memory interval, together with the estimated \ac{FDPE} $\theta^{\text{FDPE}}_{t_i}(f)$ for three characteristic time instants $t_i = \ell_i/R, \ i = \{1,2,3\}$. We can observe that both match very well, further confirming that the \ac{LO} phase walk-off is transformed into \ac{FDPN}, which manifests as an \ac{FDPE} after the \ac{LO}-\ac{PC}.

We observe that the phase error in \mbox{Fig.~\ref{fig:phase_examples}i)} is well approximated by a first-order polynomial. Since a linear phase shift corresponds to a timing offset, \ac{EEPN} manifests at $t_1$ primarily as a timing offset, which is proportional to the slope $a^{(1)}_{t_1}$ of the linear approximation. In particular, the phase changes by $\SI{1.22}{rad}$ across the signal bandwidth corresponding to a timing offset of $\SI{19.4}{\%}$ of a unit interval.

However, \ac{EEPN} can also introduce higher-order phase errors. \mbox{Fig.~\ref{fig:phase_examples}ii)} shows a quadratic phase error, approximated by a second-order polynomial, which describes a predominantly dispersive behavior proportional to $a^{(2)}_{t_2}$. The phase error in \mbox{Fig.~\ref{fig:phase_examples}iii)} requires a tenth-order polynomial for a sufficiently accurate approximation, indicating a superposition of phase errors of different orders.

Overall, this demonstrates that \ac{EEPN} manifests as a time-varying \ac{FDPE}. Consequently, the effective impairment introduced by \ac{EEPN} varies over time and may appear as a phase offset, a timing offset, dispersive behavior, higher-order \ac{FDPE}, or a combination of these.

\begin{table}[!b]
    \centering
    \caption{Receiver structures and corresponding \ac{FDPE} compensation order}
    \label{tab:mitigation_methods}
    \begin{tabular}{c@{\hspace{2.9pt}}c@{\hspace{2.9pt}}c}
        \toprule
        Receiver structure & Order $\tilde{N}$ & $\theta^{\text{comp}}_{t}(f)$ \\
        \midrule
        No compensation & -- & $\varphi_t$ \\
        Carrier phase recovery & $0$ & $a^{(0)}_{t}$ \\
        Timing recovery & $1$ & $a^{(0)}_{t} + a^{(1)}_{t} \! \cdot\! f$ \\
        Adaptive filtering & $\tilde{N}_{\text{AF}}$ & $\sum_{n=0}^{\tilde{N}_{\text{AF}}} a^{(n)}_{t} \! \cdot \! f^n$ \\
        \bottomrule
\end{tabular}
\end{table}

\section{EEPN Compensation Receiver Structures}

Building on the previous section, \ac{EEPN} can be mitigated by reversing the \ac{FDPE}. In this paper, we study the theoretical \ac{EEPN} mitigation performance of different \ac{DSP} blocks, characterized by the order $\tilde{N}$ of the \ac{FDPE} they compensate. In particular, we consider four cases (see Tab.~\ref{tab:mitigation_methods}): i)~no compensation, i.e., \ac{LO}-\ac{PC}, which removes the instantaneous \ac{LO} phase $\varphi_{t}$, ii) carrier phase recovery, which compensates a constant phase offset and corresponds to a compensation of the zeroth-order \ac{FDPE} $a^{(0)}_{t}$, iii) timing recovery, which additionally compensates linear phase components $a^{(1)}_{t}$ and corresponds to a first-order compensation, and iv) adaptive filtering, which can mitigate the \ac{FDPE} up to a given order $\tilde{N}_{\text{AF}}$.

Note that the \ac{DSP} blocks are implemented in an idealized manner. In particular, they are modeled as all-pass filters $H^{\text{comp}}_{t}(f) = \mathrm{exp} \left( -\j \theta^{\text{comp}}_{t}(f) \right)$ that reverse the \ac{FDPE} up to their respective order $\tilde{N}$, i.e., \mbox{$\theta^{\text{comp}}_{t}(f) = \sum_{n=0}^{\tilde{N}} a^{(n)}_t \! \cdot \! f^n$}, and replace the \ac{LO}-\ac{PC} block in Fig.~\ref{fig:system_model}. In practice, adaptive algorithms are required to estimate and compensate the time-varying \ac{FDPE}. To assess the maximum \ac{EEPN} compensation gain of a given \ac{DSP} block, we consider a genie-aided setting with perfectly known \ac{FDPE}, i.e., the \ac{LO} phase walk-off over time is known.

\section{Temporal Gaussian Noise Model}

Although the impact of \ac{EEPN} can be evaluated using full system simulations, its burst-like behavior requires very long simulation sequences, resulting in high computational complexity. 

Therefore, to enable fast system simulation, the \ac{TGN} model represents all system impairments, including \ac{EEPN}, as time-varying \ac{AWGN}
\vspace{-0.15cm}
\begin{equation}
    y_\ell \!= \!x_\ell + n_\ell, \ n_\ell \!\sim\! \mathcal{CN}\!\left(\!0, \sigma^2_{\text{ASE+NL+TRx}} \!+\! \sigma^2_{\text{EEPN},t=\ell/\!R}\!\right)\!.
    \label{eq:Temporal_GN_model}
\end{equation}
The term $\sigma^2_{\text{ASE+NL+TRx}}$ represents the time-invariant noise contributions from \ac{ASE}, fiber \ac{NL}, and \ac{TRx} impairments, which are neglected in this work, i.e., \mbox{$\sigma^2_{\text{ASE+NL+TRx}} = 0$}.

The key component of the model is the time-varying \ac{EEPN} distortion power
\vspace{-0.3cm}
\begin{equation}
    \sigma^2_{\text{EEPN},t} \approx \frac{1}{R} \!\!\int\limits_{-R/2}^{R/2} \!\!\left(\theta_{t}(f) - \theta^{\text{comp}}_{t}(f) \right)^2 \ \mathrm{d} f,
    \label{eq:TGN:distortion_power}
    \vspace{-0.3cm}
\end{equation}
which corresponds to the mean squared error of the residual \ac{FDPE} after the \ac{EEPN} compensation. This model significantly reduces computational complexity while still capturing the burst-like behavior of \ac{EEPN}. A reference implementation is publicly available on GitHub\footnote{Code available at \url{https://github.com/kit-cel/Temporal_Gaussian_noise_model_for_EEPN}.}.

\begin{figure*}[!t]
    \input{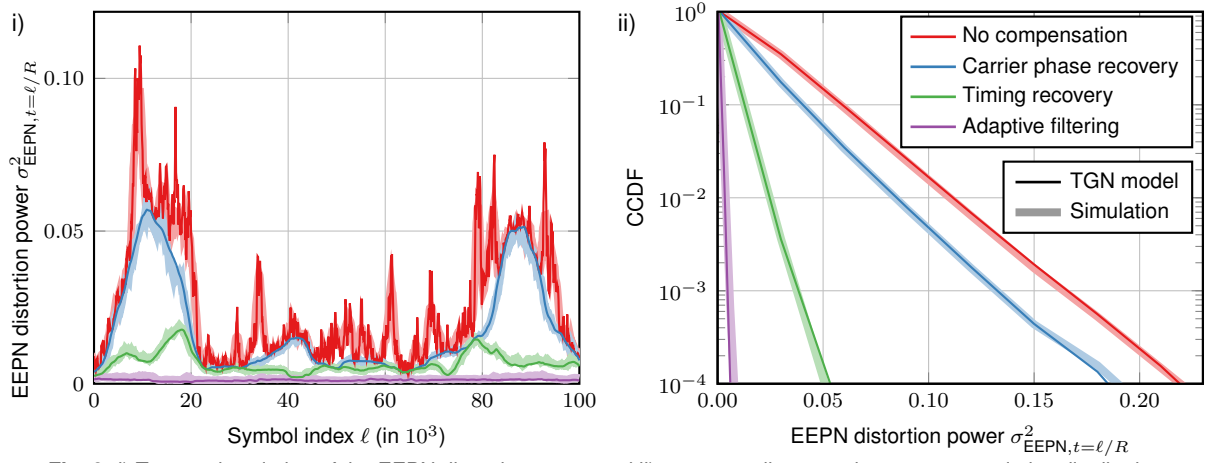}
    \vspace{-0.55cm}
    \caption{i) Temporal evolution of the \ac{EEPN} distortion power and ii) corresponding \ac{CCDF} for the comparison between simulation and the \ac{TGN} model for the \ac{DSP} blocks in Tab.~\ref{tab:mitigation_methods}.}
    \label{fig:results}
\end{figure*}

\section{Simulation Results}

In this section, we evaluate the accuracy of the \ac{TGN} model and compare the theoretical \ac{EEPN} mitigation performance of different DSP blocks. %

Fig.~\ref{fig:results}i) shows a short sequence of the \ac{EEPN} distortion powers, while Fig.~\ref{fig:results}ii) presents the corresponding statistical distribution. We can clearly observe the burst-like degradations caused by \ac{EEPN}, which appear as extended tails in the distortion power distribution. The \ac{TGN} model accurately reproduces both the temporal evolution and the corresponding distribution of the \ac{EEPN} distortion power across all compensation architectures. This shows that the \ac{TGN} model can also be used to assess post-\ac{EEPN} compensation performance.

The \ac{EEPN} distortion decreases as the \ac{FDPE} compensates increasingly higher orders. The largest improvement is obtained when moving from zeroth- to first-order compensation, i.e., when timing recovery is included. This confirms that linear phase errors (timing offsets) dominate the \ac{EEPN} distortion, as they cause the largest phase excursions. For higher-order phase compensation with sufficiently high order (here: \mbox{$\tilde{N}_{\text{AF}} = 10$}), the distortion nearly vanishes.

\section{Conclusion}

We showed that the \ac{CDC} filter transforms the \ac{LO} phase walk-off over time within the \ac{CD}-induced memory into \ac{FDPN}, which manifests as a time-varying \ac{FDPE} after the \ac{LO}-\ac{PC}. This \ac{FDPE} can appear as a phase offset, a timing offset, or higher-order phase distortions. Consequently, \ac{EEPN} can be mitigated by compensating the \ac{FDPE}, where a timing recovery yields the largest improvement.

Furthermore, the \ac{TGN} model accurately predicts the burst-like degradation caused by \ac{EEPN} through a time-varying distortion power and can also be used to model post-compensation performance. It enables accurate and low-complexity performance prediction for next-generation transmission systems affected by \ac{EEPN}.

\clearpage
\section{Acknowledgements}
This project received funding from the European Research Council (ERC) under the European Union’s Horizon 2020 research and innovation program RENEW (grant agreement No. 101001899).

\printbibliography

\end{document}